\documentstyle[multicol,aps,prl,epsf,psfig]{revtex}
\def\bi{\bibitem}
\def\be{\begin{equation}} 
\def\ee{\end{equation}} 
\def\bea{\begin{eqnarray}} 
\def\eea{\end{eqnarray}} 
\def\ba{\begin{array}} 
\def\ea{\end{array}} 
\def\bc{\begin{center}} 
\def\ec{\end{center}}

\begin{document} 
\title{Temporal and spatial correlations in a viscoelastic model
of heterogeneous faults} 
\author{Bora \" Or\c cal$^1$ and  Ay{\c s}e Erzan$^{1,2}$} 
\address{$^1$  Department of Physics, Faculty of  Sciences 
and Letters\\ 
Istanbul Technical University, Maslak 80626, Istanbul, Turkey\\ 
$^2$  G\"ursey Institute, P. O. Box 6, \c Cengelk\"oy 81220,
Istanbul, Turkey}
\date{\today} 
\maketitle 
\begin{abstract} 
We study the temporal and spatial correlations in a
one-dimensional model of a heterogeneous fault zone, in the
presence of viscoelastic effects. As a function of dynamical 
weakening and of dissipation, the system exhibits three different 
``phases" : one in which there are no time correlations between the 
events, a second, in which there are 
``Omori's law'' type temporal correlations, and a third, 
runaway phase with quasiperiodic
system size events.

PACS numbers 91.30.P, 62.20.F, 62.20.D

\end{abstract} 
\begin{multicols}{2}
\section{Introduction} 
The spatial and temporal distribution of earthquake activity has
been the foremost aim and most successful aspect of earthquake
modelling via  discerete, nonlinear networks of elastic elements
interacting via nearest neighbor couplings,
typically loaded at a constant rate far from the fault boundary. 
The Burridge-Knopoff~\cite{BK} model has been the forerunner of a
series of coarse-grained dynamical 
models~\cite{Langer,Main,Rice}, which have firmly established the
understanding of seismic activity within the paradigm of
self-organized criticality~\cite{Bak,Sornette,Olami}. These
systems exhibit ``subcritical" or ``supercritical"
deviations~\cite{Main}
from strict self-similarity due to quenched inhomogeneities,
finite driving velocities, or dissipation. 

A phenomenon which, to our knowledge, has not so far been addressed by dynamical models of the type cited above, is post-seismic relaxation~\cite{Scholz}.  It is commonly believed that viscoelastic relaxation in the immediate post-seismic period results in a redistribution of the loads, with delay times of the order of minutes, hours, or days\cite{Scholz1}. The modelling of these processes should help us understand such empirical findings as, for example, ``Omori's Law," which says that the frequency of occurrence
of ``aftershocks" decreases with time elapsed after the ``main shock" as
\be
n(t) \sim {1\over ({\rm const.} + t)^{p}}\;\;\;,\label{Omori}\ee
where $p$ is usually found 
to be very close to unity ~\cite{Main,Scholz4}.

In this paper, we mimick viscoelastic relaxation by introducing a finite
stress transfer velocity into a dynamical model recently studied by
Dahmen et al.~\cite{Dahmen}, and we investigate its effects on the
spatio-temporal behaviour of this simple model. In our coarse grained
representation, we do not claim to model the precise microscopic mechanism
for viscoelasticity, i.e., whether the relatively slow stress transfer
actually comes from multiple brittle processes in a heterogenous
medium~\cite{Scholz2} or from coupling to a viscous layer below the
litosphere~\cite{Scholz3}.  We will simply take the stress transfer
velocity ($V$) to be some effective group velocity which governs
post-event relaxation in the system and which is smaller than the velocity
of
sound~\cite{Chen}. 

The model we have used as our point of departure is
an infinitely long range (Mean Field) version of the Ben-Zion and
Rice~\cite{Rice} model which has been investigated both analytically and
numerically~\cite{Dahmen}, to reveal the presence of two different regimes
as far as spatial and temporal distributions are concerned.  It has been
found, for a narrow distribution of heterogeneities, in the limit of
infinitely slow drive, that the phase space can be described in terms of
just two parameters, the dynamical weakening $\epsilon$ and conservation
$c$, both taking values between 0 and 1. For $c<1/2$ and small $\epsilon$,
the behaviour is critical; this is the so called Gutenberg-Richter (GR)
regime, with a power law distribution of event sizes. For $c>1/2$ and
$\epsilon$ close to unity, one finds a metastable state of two-phase
coexistance, with GR behaviour interrupted by stretches of quasi-periodic,
characteristic (system-size) events, i.e. ``runaway" behaviour.

Clearly, for the 
purely Abelian models that have so far been
considered~\cite{Olami,Dahmen}, just the retardation effect coming from 
the introduction of a finite velocity of stress transfer 
cannot make any difference in the overall
statistics, since it does not
matter in which sequence the sites are updated~\cite{Dhar}.  Therefore,
the introduction of spatiality beyond that embodied in the mean
field (infinitely long range) approximation had to be considered 
simultaneously with the retardation effect. Namely, the interactions
strengths (``spring constants") were made to depend inversely on the distance, in keeping with a one-dimensional picture of the fault zone.

In this study we therefore consider simultaneously {\it i)}the effect of time
delay in the transfer of stress, {\it ii)} the decay of the
coupling strength with inverse distance, for the model of Dahmen
et al.~\cite{Dahmen} in one dimension. We have moreover considered 
a slightly different 
distribution of heterogeneities.

We find that this new model leads to three distinct phases.
The phase diagram is shown schematically in Fig.~1.
The phase space has been probed over a grid of $\Delta c=0.1$, for
$\epsilon = 0, 0.5$ and $1$.
For strongly dissipative ($c$ near 0) systems with relatively weak
``dynamical softening" effects ($\epsilon$ close to 0) we find a 
GR-like phase, with very small events which show a very steep  incipient power 
law behaviour over a rather narrow range of sizes and then cut off abruptly, and which 
display essentially no temporal correlations.  The power spectrum of the event sequence 
 is white-noise.
For intermediate values of these parameters, the event distribution is similar 
to the first, however the power spectrum  reveals non-trivial 
temporal correlations, a feature not observed by Dahmen et al.~\cite{Dahmen} in the 
GR phase.
In the region of  $c$ close to unity (strong conservation)we find 
quasiperiodic runaway behaviour.
\begin{figure}
\begin{center}
\leavevmode
\psfig{figure=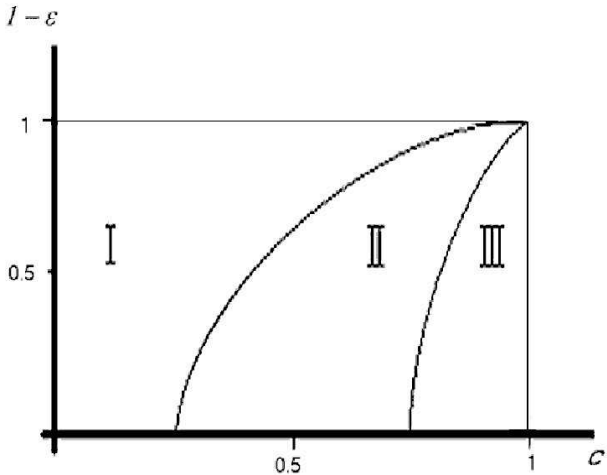,width=8cm,height=6.4cm,angle=-90}
\end{center}
\narrowtext
\caption{
Schematic phase diagram exhibiting three different ``phases''
where, I.) small events delta correlated in time, II.) small events with
Cauchy type time correlations, and III.) quasiperiodic, system-size
runaway events, dominate.}
\end{figure}

The phenomenology of each of these phases is rather rich.
In regions of  strong
heterogeneity, one may  observe patches of blocks to slip in 
unison, pinned at
either end by relatively large threshold stresses, exhibiting
quasiperiodic behaviour within a sea of power law events.  
We have not observed  the switching behaviour between 
coexisting GR and runaway phases, as reported for the mean field
model~\cite{Dahmen}, but this may be because 
this would require prohibitively long simulations in our 
scheme. The distribution of the accumulated stresses along the fault 
zone is both qualitatively and quantitatively similar in all the three 
regions, in contrast to the previous findings.

It should be stressed here that our approach is a departure from the usual
quest for scale invariant spatial or temporal distributions. With the
introduction of a finite stress transfer velocity (which we take to be
unity), and a finite driving velocity, we in fact have three well
seperated time scales in the problem: That of the driving velocity (the
largest time scale), the viscoelastic time scale, and the triggering time
scale (where slip occurs instantaneously).

The paper is organized as follows. In section 2, the precise
definiton of the model is given.  
In section 3, we report our results for the statistics of the 
magnitudes integrated over time scales corresponding to typical
event durations, which should be compared with those in the 
infinite stress transfer velocity/zero driving velocity limit. 
We then go on to compute temporal and spatial  correlation functions for 
coarse grained events.
In section 4 we provide a discussion of our findings.

\section{The Model}

We consider a one-dimensional  array of finite segments, or
blocks. The local stress $\tau_i$ on the $i$th block is given, at
time $t$ by
\be
\tau_i(t)=\sum_{r=-R}^{R} k_r [u_{i+r}(t-r/V)-u_i(t)] +K[vt-
u_i(t)]\label{tau1}
\ee
where the range of the interaction, $R$, is of the
order of the system size, $u_i(t)$ is the offset of the $i$th
block in the direction of the constant driving velocity $v$, at
time $t$; $K$ is the effective shear modulus, and $k_r = k/\vert
r\vert$ is the elastic coupling between blocks seperated by a distance $r$. As long as all
the $\tau_i<\tau_{s,i}$, where $\{\tau_{s,i}\}$ are randomly
distributed failure stresses, the system is immobile. 

The viscoelastic stress relaxation is mimicked by the delay, $r/V$, in the
transfer of stress. We shall henceforth set $V$, the velocity for the
stress transfer along the blocks, to unity. 
Note that $V$ is not typically the sound velocity, but
some effective group velocity smaller than that of sound, governing the
processes of viscoelastic stress relaxation in this coarse grained
model.~\cite{Scholz,Chen}

The dynamics is defined as follows.
If the threshold value is exceeded  at some $i$, at time $t$,
then 

{\it i)}The stress at the $i$th block is reduced by 
\be 
\delta \tau_i=\tau_{s,i}-\tau_{a,i}\;\;,
\ee 
where the $\{\tau_{a,i}\}$ are  random arrest
stresses. 

{\it ii)} The value of the failure stress at the  $i$'th block 
is reset, until all motion once more ceases, to a ``dynamical''
threshold value 
\be \tau_{d,i}=\tau_{s,i}-\epsilon (\tau_{s,i}-\tau_{a,i})\;\;,
\ee 
where $\epsilon$  parameterizes the dynamical weakening effect.

{\it iii)} The stress drop is redistributed, according to
Eq.(\ref{tau1}), so that  $\tau_{i+r}$ is incremented, at the
$t+\vert r\vert $'th time step by 
\be
\delta \tau_{i+r}= c_r(\tau_{s,i}-\tau_{a,i}) \;\;\;\;\;\;\;\;\;\;\;\;\;
 c_r={k_r\over \sum_{r^\prime}  k_{r^\prime} +K }
\ee
We may once more define $c\equiv\sum_r c_r$, with  
$0\le c\le 1$,  to be
the parameter which measures the degree of conservativeness of
the system, although it should be noted that this definition now
involves an implicit integral over time as well as space.

The boundary conditions are fixed, so that $u_1=u_L\equiv 0$.
At each time step, the stress at all the blocks $i$ is
recalculated according to Eq.(\ref{tau1}). This means that the
constant drive term $Kvt$ is incremented also.

\section{Simulations}

Since the finite stress transfer velocity introduces a definite time scale
into the system, which also sets the characteristic time scale of the
event duration, one now faces the problem of having to go to extremely
long runs with a driving velocity which is at least six to seven orders
\begin{figure}
\begin{center}
\leavevmode
\psfig{figure=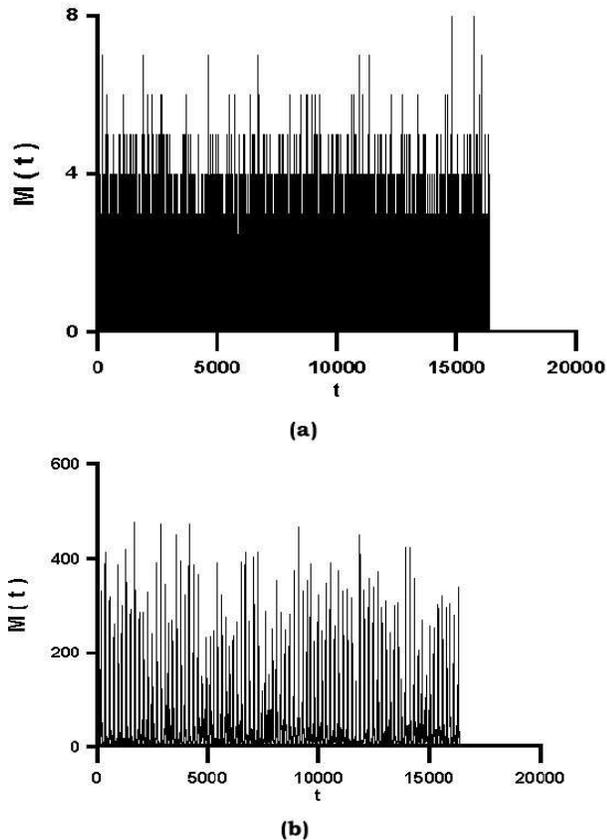,width=8cm,height=11.2cm,angle=0}
\end{center}
\narrowtext
\caption{
Typical time series for the magnitudes $M$ v.s. time, for
16384$\times128$ time steps.  
a) in the ``Gutenberg-Richter'' (small event) and b) the runaway regimes.
Note the difference in the vertical scales.  Shown are plots for a)
$c=0.16$, $\epsilon=0.5$; b) $c=0.9$, $\epsilon=0.5$.  The driving
velocity is $v=10^{-5}$ and the shear modulus $K=1$ for this and the
following figures.}
\end{figure}
\noindent 
of magnitude smaller than the latter.  The zero driving velocity
trick of
simply scanning the system for that site which is closest to slipping and
loading all sites by the missing amount, is no longer appropriate here -
one has to first check that there are no stress ``parcels" still on the
way. The actual simulation times get prohibitively large as a result, and
we had to be content with large but finite ranges of interaction, up to
1/6 the fault size, and with a one-dimensional fault.

{\bf Table I.}
\narrowtext{
Values of the Gutenberg-Richter exponent $b$ in the different regimes.}
\bc 
\begin{tabular}{|c|c|c|c|} 
\hline 
Region &~~~~~$c$~~~~~ &~~~~~$\epsilon$~~~~~ &~~~~~$b$~~~~~\\
\hline\hline 
I &0.15&       0.5 &       5.2  \\ \hline 
II& 0.65&       0.5 &       5.2  \\ \hline 
III& 0.9 &       0.5 &       1.6   \\ \hline 
\end{tabular}   
\ec 

We have simulated the system described in the preceding section
on a grid of 300 blocks, with the range of interactions going up to 
$R=50$. The distribution of
stress drops $\delta \tau_i =\tau_{s,i}-\tau_{a,i}$
was chosen to have the form
$p(x)\propto x^{-\mu}$, with $\mu=1.2$ However, upon finding that 
arbitrarily large stress drops pinned the edges of finite segments 
in the fault and distorted the distribution of event sizes, we decided to 
limit the range of the $\delta \tau$ to a width comparable to those considered in 
Ref.~(~\cite{Dahmen}), namely 0.2. It is generally found that long active 
fault zones organize themselves into states with relatively small 
heterogeneity, and our results should be considered in this spirit. 

The driving velocity $v$ is taken to be $10^{-5} V$ in the simulations
reported below. It should be noted that larger
driving velocities result in individual cells exceeding their threshold
values and collapsing independently from their neighbors, and as a result
the system never achieving a self organized state. With realistic driving
velocities of $\sim 10^{-9}$ m/sec, $v=10^{-5} V$ corresponds to a stress
transfer velocity of $\sim 10^{-4}$ m/sec.  For ``block sizes" of $\sim
10^2$ meters setting our lattice spacing, a stress transfer velocity of
$V\sim 10^{-4}$m/s corresponds to time steps of duration $10^6$ s.

In Figure 2, we display the typical time series resulting from
plotting the integrated magnitudes $M(t)$ for different values of 
the system parameters. These are obtained by
summing over the number of blocks where the threshold has been
exceeded within an interval $\delta t=128$.  
\begin{figure}
\begin{center}
\leavevmode
\psfig{figure=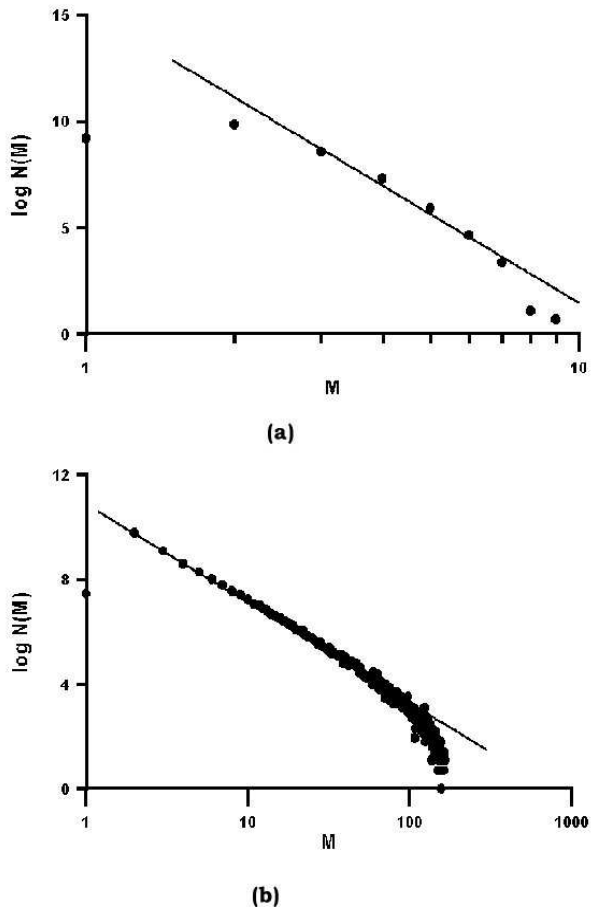,width=8cm,height=12cm,angle=0}
\end{center}
\narrowtext
\caption{
The frequency v.s. magnitude plots on a double
logarithmic scale, for the same set of parameters as a) and b) in
Figure 2. The lines are intended as a guide to the eye with 
slopes $-5.2$ and $-1.6$, respectively.}
\end{figure}
This value was chosen 
as approximately that time interval needed for a signal originating in the
middle 
of the fault to be able to reach the edges of the fault. This time-coarse- 
grained way of identifying events is in keeping with the way
earthquake data is taken, with the time integral taken over the 
actual displacements of the seismographs.
Thus, we define,
\be
M(t)=\sum_{t^\prime= t}^{t+\delta t}\sum_i [\Delta u_i(t^\prime)]^0\;\;\;, \label{M}
\ee
where the zeroth power of the slip $\Delta u_i(t^\prime)$ is taken so that 
$M(t)$ simply 
counts the number of slipped blocks within the time interval
$\delta t$.  
The time series $M(t)$ reveal 
no immediately observable differences between the regions
I and II shown in Fig.1; therefore we have selected only one set 
of parameter values to illustrate both these regions (see Fig. 2a).
On the other hand the much larger 
magnitudes observed in the runaway region, and 
their marked quasiperiodicity are apparent 
in Fig. 2b.

The regions I and II are also similar in the way the frequency $f(M)$ of
events scales with the magnitude $M$, for a given binning size $\delta t$.  
In Figure 3, we show the plots of the frequency $f(M)$ v.s. magnitude $M$  in the
small event (the ``Gutenberg-Richter" phase found in Ref.~\cite{Dahmen}) and 
in the
runaway regime, for the same parameter values as in Fig. 2. The ``power
law" fits, $f(M) \sim M^{-b}$, to the small event regime (regions I and
II) are poor, and can only be thought of as suggestive; they extend over
too small a range to really signify self--similarity.  Notice that in the
``runaway" regime (region III), the magnitudes cover a wider
range; however we do not
observe as marked a ``super-criticality," i.e., a frequency of {\it large} events in excess of a power law size distribution, as has been
reported elsewhere~\cite{Main,Rice,Anghel}. 

\begin{figure}
\begin{center}
\leavevmode
\psfig{figure=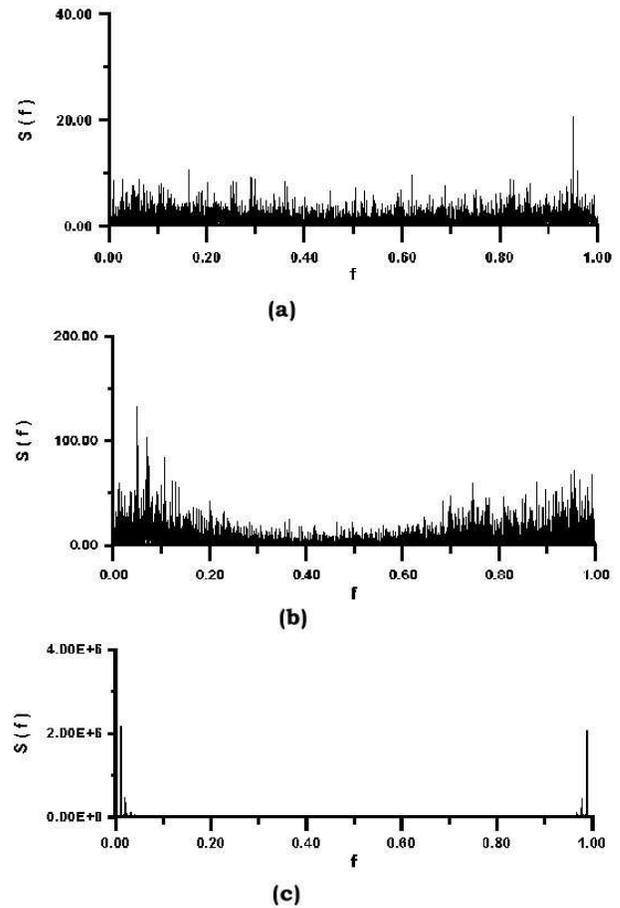,width=8cm,height=12cm,angle=0}
\end{center}
\narrowtext
\caption{
Power spectra of the time series of magnitudes, computed for parameter
values in the regions I, II, III of the phase diagram shown in Fig.(1).
a) $
c=0.16$, $\epsilon=0.5$, b) $ c=0.72 $, $\epsilon=0.5$, c)  $ c=0.89$,
$\epsilon=0.5$.
The data for panels a) and c) are taken over series of 8192$\times$128
time steps. The data in panel (b) have been averaged over 7 runs of
16384$\times$128 time steps.}          
\end{figure}

Within regions I and II, the Gutenberg-Richter exponent $b$ is 
 sensitive to the driving velocity $v$ and also depends,
less strongly, on the parameters $c$, $\epsilon$. Here the distribution is
very steep, with $b$ ranging between 4 and 5. The values obtained are
given in Table I.  The ``runaway'' regime exhibits much more realistic $b$
values, around 1.5 to 1.6.
One should note, moreover,  that since the system is no longer scale 
invariant,  the statistics of the magnitudes $M$ 
are also sensitive to the binning size $\delta t$, so that the $b$ values
here are only useful for purposes of discriminating between different
regions of the phase space.

The interesting difference between the three regimes delienated in Fig. 1 
become apparent in the power spectra, 
\be 
S(f)=\int dt e^{i 2\pi f t} C(t)\;\;\;,
\ee 
where 
\be C(t) ={1 \over T}\int_0^T dt^\prime M(t^\prime) M(t^\prime+t)\;\;\label{Correl}\ee
is the time-correlation function for the
coarse grained 
magnitudes $M(t)$, where $t$ stands for the number of time intervals
$\delta t$, for a total time of measurement extending over a period $T$.

\begin{figure}
\begin{center}
\leavevmode
\psfig{figure=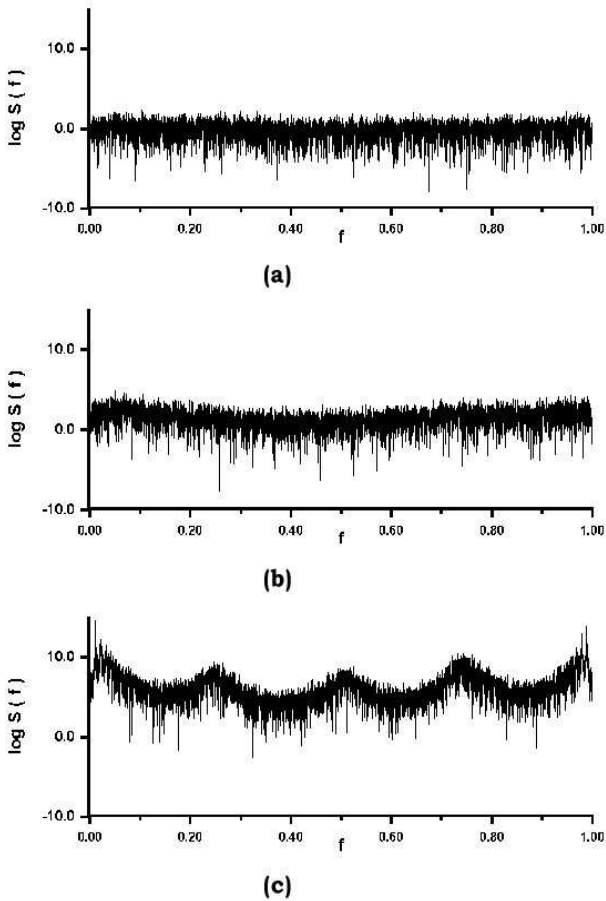,width=8cm,height=12cm,angle=0}
\end{center}
\narrowtext
\caption{
 The same power spectra as in Fig. 4, on a logarithmic scale.
Panels a), b), and c) belong to the regions I, II, III of the phase
diagram in Fig. 1.}
\end{figure}

In Figures 4 and 5 we display linear and semi-logarithmic plots of the power 
spectrum in the three different phases.
It can be seen in Fig.(4a) and in a more pronounced 
way in the logarithmic plot of the same power spectrum 
in Fig.(5a) that in the 
``small event"
regime I (panel a), the power spectrum is essentially flat, white-noise like, 
indicating an absence of correlations between the earthquakes, i.e., 
$C(t)\propto \delta (t)$.  
For intermediate values of $c$ and $\epsilon$, i.e., in the small-event region II, however,  
we find that superposed upon the white-noise like background, the 
upper envelope of the power spectrum displays a distinctive curve, (see
Fig.4b)
indicating the presence of non-trivial temporal correlations. 
In the runaway region (region III), we find a markedly different,
quasiperiodic
behaviour, as can be see from Figs. (4c) and (5c).  The very pronounced
peak in the power spectrum near the origin is large enough to suppress all
the others; we can see the other frequencies that are present only in the
logarithmic plot.

In Figure 6, we show the result of taking an inverse transform of the 
{\em envelope} (roughly the highest points) of the power spectrum shown in
Fig. 4b.  This crude estimate of
the time correlation function is corraborated by a
more careful evaluation, to which we now turn.

\begin{figure}
\begin{center}
\leavevmode
\psfig{figure=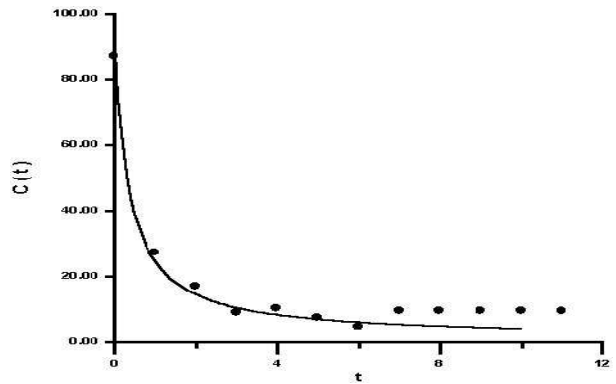,width=8cm,height=5cm,angle=0}
\end{center}
\narrowtext
\caption{A crude estimate for the time correlation function, from 
the inverse Fourier transform of the points in the envelope of
the power spectrum, shown in Fig. 4b.  The fit is to $1+85(1+3t)^{-1}$.}
\end{figure}

We have computed  $C(t)$ directly
from a time series of 81920 time intervals (of 128 steps each). 
We have normalized the correlation function  by 
$(1/T)\int_0^T M^2(t^\prime)dt^\prime$, so
that $C(0)=1$.  We find that the normalised $C(t)$ can be fit rather
well by a function of the form  
\be
C(t) = A + {B \over D + t}\;\;,\label{Omori1}\ee

\begin{figure}
\begin{center}
\leavevmode
\psfig{figure=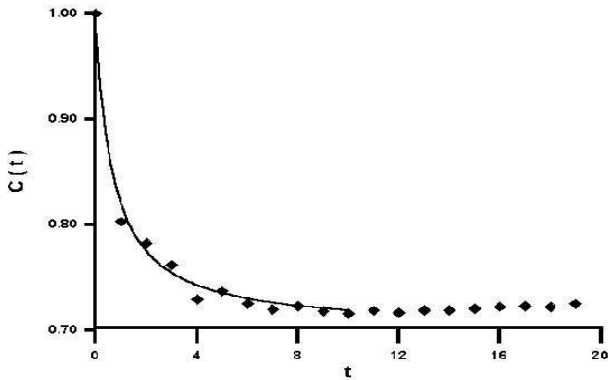,width=8cm,height=5cm,angle=0}
\end{center}
\narrowtext
\caption{
The time correlation function $C(t)$ in region II of the phase
diagram ($c=0.65,\;\; \epsilon=0.5$), computed from Eq.(8), over a 
time series of
81920 $\times$ 128 steps. The fit is to $0.7+0.2(0.66+\,t)^{-1}$.}
\end{figure}
\noindent
with $A=0.7,\; B=1/5,\; D=2/3$. 
Our results are shown in Fig.7.

Note that the time correlation function $C(t)$ measures the average
frequency with which a time lapse $t$ seperates two events, {\em weighted}
by the magnitude of the events. In another way of saying the same thing,
it is the weighted average of the number of times that one registers pairs
of events seperated by a time lapse equal to $t$. If there is no event
taking place at time $t+t^\prime$ after an event at time $t^\prime$, i.e.,
if $M(t^\prime) \neq 0$, but $M(t+t^\prime)=0$, there will be no
contribution to the integral for $C(t)$ in (\ref{Correl}).

We would like to recall, at this point, Omori's Law (\ref{Omori}) for the
frequency of aftershocks~\cite{Main,Scholz4}. Actually, geophysicists are
generally hesitant to label a given shock as either an ``aftershock" or
``main schock," and admit that these are conventional distinctions, which
are difficult to make precise in a quantitative way.  Viewed in this way,
Omori's law is just a statement of the relative frequency of pairs of
events seperated by a time $t$, and is a slightly cruder version of the
time correlation function. We see that the form we find for the time-decay 
of the 
correlation function matches that of Omori's Law, with a power $p=1$, as
found for real earthquake statistics.

From Fig. 7, and Eq.(9), we see that $C(t)-A$
drops by a factor of 1/2 within one time interval (consisting of 128 time
steps). 
Since we have already estimated the time steps here 
to correspond to about $10^6$ seconds, this means correlations times of the 
order of $10^8$ s, namely $\sim 3$ yrs, which is quite realistic for the 
time interval in which aftershocks die away after a big event.
\begin{figure}
\begin{center}
\leavevmode
\psfig{figure=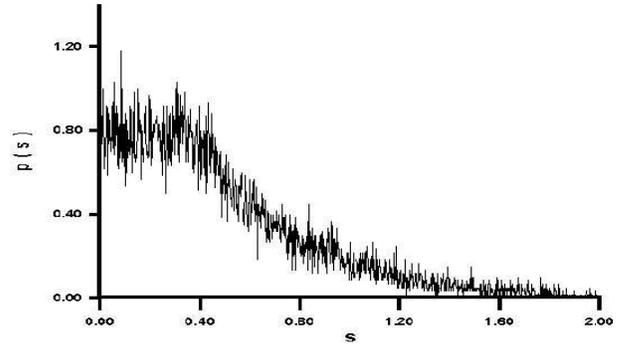,width=8cm,height=4.5cm,angle=0}
\end{center}
\narrowtext
\caption{
The probability density of the fraction $s$ of the slipping
stress (see text) computed along the fault zone, for $c=0.89$,
$\epsilon=0.5$ (region III). 
The histogram is averaged over 20 snapshots, seperated by $5\times 10^4$
timesteps. The figures for regions I and II are indistinguishable from
this one.}
\end{figure}

To investigate the spatial correlations in the system, we first considered 
the distribution of the fraction of the slipping stress on each block,
\be
s_i\equiv {\tau_{s,i}-\tau_{i}\over \langle \tau_{s,i}-\tau_{a,i}\rangle }
\;\;\;, \label{s}
\ee
along the fault zone. 
This is a quantity in which one might have expected 
 a greater self-organization building up as one goes from region I to
region III in the phase diagram, yet we did not find this to be the case.
In contrast to Ref.~\cite{Dahmen}, in the present model the distribution
$p(s)$ remains essentially invariant in all the three regions, and has the
shape shown in Fig. 8.  Similarly, we found that the equal time spatial
correlations $\langle s_i s_{i+r}\rangle$ between the fraction of the
slipping stresses accumulated at
each site, showed essentially delta function behaviour in all three
regions, with the correlations never extending beyond next nearest
neighbors.  On the other hand, defining the coarse grained toppling  
variables 
\be m_c(t, i) \equiv\sum_{t^\prime=t}^{t+\delta t}[ \Delta
u_i(t^\prime)]^0\;\;\;\; , \label{coarse}
\ee 
and setting   $m_c(i)=0$ at sites beyond the
boundaries of the
fault  we found
that the correlation function 
\be C_m(r)={\langle m_c(t,i) m_c(t,i+r)\rangle - 
\langle m_c(t,i)\rangle^2 \over \langle
m_c(t,i)\rangle^2}\;\;\;,\label{correl} 
\ee 
where the averages are
performed both over $i$ and $t$, indeed displayed markedly different
behaviour in region III, in comparison to I and II. Our results are shown
in Fig. 9. The
time averages were performed over 60 snapshots, taken at intervals of
$5\times 10^4$ timesteps.
The correlations are negligible (of the order of $10^{-4}$) in
the first two regions, whereas, in region III, one sees a gradual decay.  
A straight line fit to the semilogarithmic plot (Fig. 10) suggests an 
exponential decay and gives a
correlation length
of 30 lattice units, corresponding to $\sim 3\times 10^3$ meters.

\begin{figure}
\begin{center}
\leavevmode
\psfig{figure=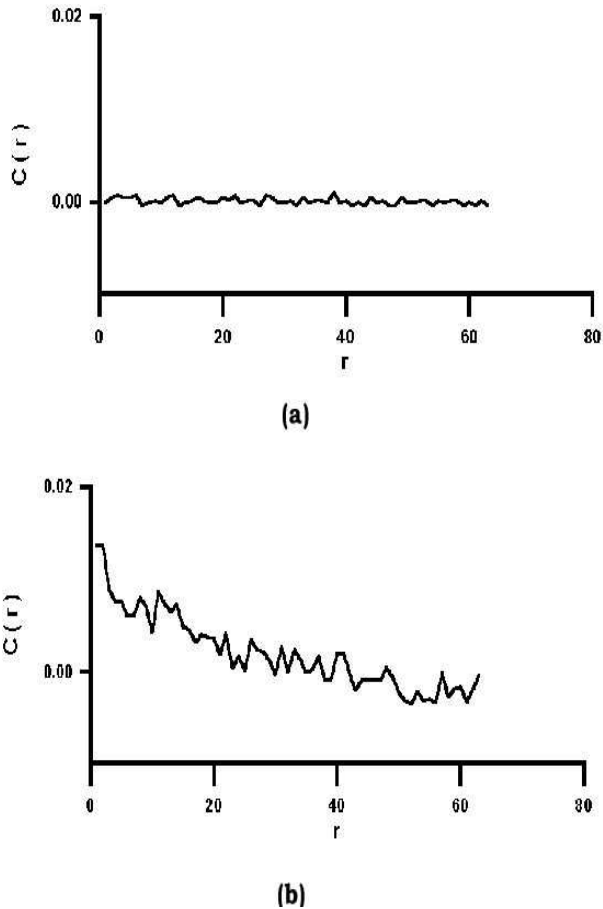,width=8cm,height=12cm,angle=0}
\end{center}
\narrowtext
\caption{The spatial correlations between toppling events coarse grained
in time, 
in the regions I and III of the
phase diagram. a) $ c=0.16$, $\epsilon=0.5$, b)
 $ c=0.89$, $\epsilon=0.5$. 
The plot for  $ c=0.54 $,$\epsilon=0.5$ (in region II)
looks identical to panel a).
 Averages have been taken
over 60 snapshots seperated by time intervals of 50,000 steps.}
\end{figure}

\section{Discussion}

The inclusion of viscoleastic effects into the study of crack
propagation
 and pinned driven systems has recently made important progress
~\cite{Kessler,Marchetti}, and promises to be fruitful also in the
modeling of earthquakes. The most important outcome of introducing
viscoelastic effects is to be found in the more subtle temporal
correlations between events; for highly dissipative systems the
correlations are delta function like, whereas for intermediate values of
the dissipation, one observes correlations that decay as $\sim ({\rm const.} + t)^{-1}$ between events, which is of the form of Omori's Law (\ref{Omori}).
To our knowledge, this is the first demonstration of how Omori's Law may
arise in such a system. 

Our preliminary findings
indicate that, due to viscoelastic effects, the runaway phase (region III)
of quasiperiodic events in the present model of a heterogeneous fault zone
is pushed to a relatively smaller region of the phase diagram~(see Fig. 1)
than found previously~\cite{Dahmen} and the frequency distribution in this
phase displays scale invariance over a sizable region of event sizes.  
We have verified that 
this region is distinguished by relatively long range spatial correlations
between slipping events, in contrast to the ``small event" regimes.

\begin{figure}
\begin{center}
\leavevmode
\psfig{figure=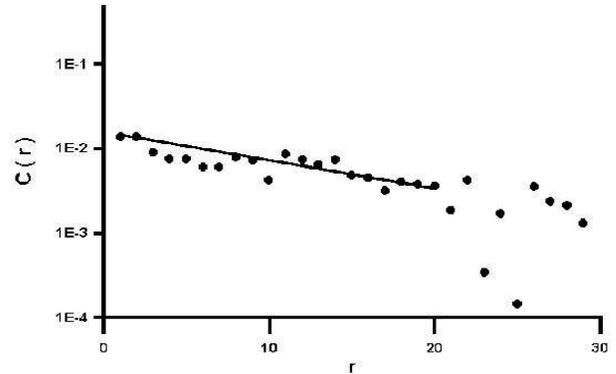,width=8cm,height=5cm,angle=0}
\end{center}
\narrowtext
\caption{
Semilogarithmic plot of the spatial correlations in region III.  The
slope of the straight line fit gives a correlation length of 30
lattice units, 
or equivalently, $3\times 10^3$ meters.} 
\end{figure}

It has been remarked before\cite{Main,Anghel} that various
system-dependent features, notably dynamical weakening and dissipation,
introduce time and length scales into the problem and take the system away
from criticality.  We would like to remark that one may reverse the
emphasis here to say that rather than the scaling region of the
Gutenberg-Richter regime, one should examine the sub--or super--critical
behaviour to characterise a specific fault zone. In particular, we have
shown that determining the nature of the space  and time-correlations in
the system gives important clues as to the relative degree of dissipation
or dynamical weakening.

{\bf Acknowledgements} 

It is great pleasure to thank Mustafa Aktar,  Deniz Erta\c s  and
Celal \c Seng\"or 
for several useful discussions and for making available to us a
number of references. One of us (AE) gratefully acknowledges
partial support from the Turkish Academy of Sciences.

\end{multicols}

\begin{thebibliography}{26} 
\bi{BK} R. Burridge and L. Knopoff, Bull.
Seismo. Soc. Am., {\bf 57}, 341 (1976). 
\bibitem{Langer} J.M. Carlson,
J.S. Langer, B.E. Shaw, 
Rev. Mod. Phys.
{\bf 66}, 657 (1994). 
\bi{Main} I. Main, Reviews of Geophysics, {\bf 34},
433 (1996). 
\bi{Rice} Y. Ben-Zion and J.R. Rice, J. Geophys. Res. {\bf
98}, 14109 (1993); J. Geophys. Res. {\bf 100}, 12 959(1995); J. Geophys.
Res. {\bf 101}, 5677 (1996). 
\bi{Bak}P. Bak and C. Tang, J. Geophys. Res. {\bf 94}, 15635 (1989). 
\bi{Sornette}D. Sornette and A. Sornette,
Europhys. Lett. {\bf 9}, 197 (1989). 
\bi{Olami} Z. Olami, H.J.S. Feder,
and K. Christensen, Phys. Rev. Lett., {\bf 68}, 1244 (1992). 
\bi{Scholz} C.H. Scholz, {\it The Mechanics of Earhquakes and Faulting}
(Cambridge University Press, Cambridge, 1990).
\bi{Scholz1} Ibid., p. 314ff.
\bi{Scholz4} Ibid., p. 205.
\bi{Scholz2} Ibid., p. 34.
\bi{Scholz3} Ibid., p. 236.
\bi{Dahmen} K. Dahmen, D. Erta\c s, and Y. Ben-Zion, Phys. Re. E {\bf58},
1494 (1998).
\bi{Chen} Y.T. Chen and L.
Knopoff, ``The quasistatic extension of shear crack in a viscoelastic
medium," Geophys. J. R. astr.  Soc. {\bf 87}, 1025 (1986). 
\bi{Dhar} D. Dhar, Phys. Rev. Lett. {\bf 64}, 1613 (1990).
\bi{Kessler}
D.A. Kessler and Levine, Phys. Re. E {\bf 59}, 5154 (1998).
\bi{Marchetti} M.C. Marchetti, A.A. Middleton, T. Prellberg,
``Viscoelastic depinning of driven systems: mean field plastic scallops,"
cond-mat/9912461.
\bi{Anghel} M.
Anghel, W. Klein, J.B. Rundle, J. S. S\'a Martins, ``Scaling in a Cellular
Automaton Model of Earthquake Faults," cond-mat/0002459. 
\end{thebibliography}
\end{document}